%% file: 000_MAIN.tex
\documentclass[sigconf,authordraft=false, review=false]{acmart}

\usepackage{svg}

\usepackage{xcolor}
\newif\ifdraft
\drafttrue
\ifdraft
\newcommand{\jsnote}[1]{\textcolor{red}{***Johannes: #1}}
\newcommand{\gmnote}[1]{\textcolor{blue}{***Gonzalo: #1}}
\newcommand{\mbnote}[1]{\textcolor{green}{***Matthias: #1}}
\else
\newcommand{\jsnote}[1]{}
\newcommand{\gmnote}[1]{}
\newcommand{\mbnote}[1]{}
\fi

\usepackage{placeins}

\usepackage{listings}
\definecolor{verylightgray}{rgb}{.97,.97,.97}
\definecolor{darkgreen}{rgb}{.0,.235,.0}

\lstdefinelanguage{circom}{
	keywords=[1]{assert, do, else, false, function, if, length, public, true, while, for}, 
	keywordstyle=[1]\color{blue}\bfseries,
	keywords=[2]{var, signal},	
	keywordstyle=[2]\color{teal}\bfseries,
	keywords=[3]{include, template, input, output, private, main, component},	
	keywordstyle=[3]\color{violet}\bfseries,
	identifierstyle=\color{black},
	sensitive=false,
	comment=[l]{//},
	morecomment=[s]{/*}{*/},
	commentstyle=\color{gray}\ttfamily,
	stringstyle=\color{darkgreen}\ttfamily,
	morestring=[b]',
	morestring=[b]"
}
\usepackage{amsmath}

\newtheorem{definition}{Definition}

\usepackage[vlined,linesnumbered]{algorithm2e}
\makeatletter

\usepackage[nolist]{acronym}
\acrodefplural{SNARK}[SNARKs]{succinct non-interactive arguments of knowledge}
\acrodefplural{STARK}[STARKs]{scalable transparent arguments of knowledge}
\begin{acronym}
\acro{CDF}[CDF]{cumulative distribution function}
\acro{DP}[DP]{differential privacy}
\acro{LDP}[LDP]{local differential privacy}
\acro{GDPR}[GDPR]{general data protection regulation}
\acro{PDF}[PDF]{probability density function}
\acro{QAP}[QAP]{quadratic arithmetic program}
\acro{SNARK}[SNARK]{succinct non-interactive argument of knowledge}
\acro{STARK}[STARK]{scalable transparent argument of knowledge}
\acro{TGM}[TGM]{truncated geometric mechanism}
\acro{ZKP}[ZKP]{zero-knowledge proof}
\acro{R1CS}[R1CS]{rank one constraint system}
\end{acronym}

\AtBeginDocument{%
  \providecommand\BibTeX{{%
    \normalfont B\kern-0.5em{\scshape i\kern-0.25em b}\kern-0.8em\TeX}}}

\copyrightyear{2022} 
\acmYear{2022} 
\setcopyright{acmcopyright}\acmConference[ARES 2022]{The 17th International Conference on Availability, Reliability and Security}{August 23--26, 2022}{Vienna, Austria}
\acmBooktitle{The 17th International Conference on Availability, Reliability and Security (ARES 2022), August 23--26, 2022, Vienna, Austria}
\acmPrice{15.00}
\acmDOI{10.1145/3538969.3538992}
\acmISBN{978-1-4503-9670-7/22/08}

\begin{document}

\title{Towards Verifiable Differentially-Private Polling}

\author{Gonzalo Munilla Garrido}
\authornotemark[1]\thanks{*Corresponding author.}
\email{gonzalo.munilla-garrido@tum.de}
\orcid{0000-0002-0135-9432}
\affiliation{%
  \institution{Technical University of Munich}
  \streetaddress{Boltzmannstr. 3}
  \city{Garching}
  \country{Germany}
  \postcode{85748}
}
 
\author{Matthias Babel}
\email{matthias.babel@fim-rc.de}
\orcid{0000-0002-8794-2126}
\affiliation{%
  \institution{FIM Research Center, \\University of Bayreuth}
  \streetaddress{Wittelsbacherring 10}
  \city{Bayreuth}
  \country{Germany}
  \postcode{95447}
}

\author{Johannes Sedlmeir}
\email{johannes.sedlmeir@fim-rc.de}
\orcid{0000-0003-2631-8749}
\affiliation{%
  \institution{Fraunhofer FIT, Branch Business \& Information Systems Engineering}
  \streetaddress{Wittelsbacherring 10}
  \city{Bayreuth}
  \country{Germany}
  \postcode{95447}
}

\begin{abstract}
  Analyses that fulfill differential privacy provide plausible deniability to individuals while allowing analysts to extract insights from data. However, beyond an often acceptable accuracy tradeoff, these statistical disclosure techniques generally inhibit the verifiability of the provided information, as one cannot check the correctness of the participants' truthful information, the differentially private mechanism, or the unbiased random number generation. While related work has already discussed this opportunity, an efficient implementation with a precise bound on errors and corresponding proofs of the differential privacy property is so far missing. In this paper, we follow an approach based on zero-knowledge proofs~(ZKPs), in specific succinct non-interactive arguments of knowledge, as a verifiable computation technique to prove the correctness of a differentially private query output. In particular, we ensure the guarantees of differential privacy hold despite the limitations of ZKPs that operate on finite fields and have limited branching capabilities. We demonstrate that our approach has practical performance and discuss how practitioners could employ our primitives to verifiably query individuals' age from their digitally signed ID card in a differentially private manner.
\end{abstract}

\ccsdesc[300]{Information systems~Electronic data interchange}
\ccsdesc[500]{Security and privacy~Cryptography}
\ccsdesc[300]{Security and privacy~Human and societal aspects of security and privacy}
\ccsdesc[500]{Security and privacy~Privacy-preserving protocols}

\keywords{Digital wallet, exponential noise, privacy, randomized response, SNARK, survey, zero-knowledge proof}

\maketitle

\input{001_Introduction}

\input{002_Preliminaries}
\input{003_Implementation}
\input{004_Evaluation}

\input{005_Related_work}

\input{006_Discussion}

\input{007_Conclusion}

\begin{acks}
We thank the Bavarian Ministry of Economic Affairs, Regional Development and Energy for their funding of the project ``Fraunhofer Blockchain Center (20-3066-2-6-14)'', and the Ethereum foundation's grant that made this paper possible. 
\end{acks}

\bibliographystyle{ACM-Reference-Format}
\bibliography{999_REFS}


\end{document}

%% file: 001_Introduction.tex
\section{Introduction}
\label{sec:Introduction}

Gathering information through polls to produce statistics regarding, e.g., the health, financial status, or demographics of a population bears the risk of exposing individuals' sensitive data during and after the survey. One approach is to anonymize the gathered data centrally, which implies high costs for implementing security measures and still carries ethical risks. Moreover, since interviewees cannot control that their data is adequately anonymized and protected in this paradigm and their level of trust in the surveyor is limited, their response may be subject to bias, specifically with highly sensitive or embarrassing questions.

A simple means to enhance privacy by design and reduce the risk of bias in such polls is through randomized response~\citep{warner1965randomized}, its variations~\cite{RR_two_stage_1, RR_two_stage_2, RR_question_model, RR_forced_response, yang_wireless_2016}, or more involved forms of local differential privacy (DP)~\cite{RR_local_DP_exterme_points, RR_local_DP_exterme_points_graphs, RR_DP_extremal, heavy_hitter_agg, binary_mech, DP_precision_Geometric}, which provide plausible deniability by adding noise to interviewees' answers.
The noise distribution of these techniques is typically centered around $0$ with finite variance~\cite{dwork_algorithmic_2013}, so according to the law of large numbers, the mean of the noisy data converges towards the original mean as the sample size increases, improving accuracy.
However, in this local approach, there is a lack of response \emph{verifiability} -- the interviewer has no assurance that the interviewee, i.e., the adversary in our system, answered truthfully.
This lack of verifiability is arguably severe when rewards encourage malicious participation, e.g., there is a monetary incentive to participate but no willingness to answer truthfully.

Verifiable computation can prove the execution of a particular algorithm from truthful inputs without revealing private information~\cite{ben2019scalable}. 
Accordingly, we suggest combining verifiable computation with \ac{LDP} techniques to prove (i)~the interviewees' plausible deniability guarantee derived from randomness and (ii)~the truthfulness of their deterministic answer, i.e., the value has been signed by a reputed authority. 
Similar approaches have been discussed, for instance, in~\cite{narayan2015verifiable,rueckel2022fairness}.
Our approach hence targets polls where there is cryptographic evidence for the answers, e.g., a digital ID card signed by a government, digital diplomas issued by a certified university, or COVID-19 immunity passports certified by pharmacies or doctors. 
Such attestations are considered, for instance, in the European digital wallet initiative~\cite{eu2021digitalwallet,rieger2022not,sartor2022ux}. In this context, it is particularly helpful that the digital certificates involved in the many implementations of digital wallets are in fact anonymous credentials~\cite{camenisch2001efficient,schlatt2021kyc, sedlmeir2021identities}, which allows us to extract a user's attribute values without revealing strongly correlating information. Moreover, we believe that in the private sector, attestations derived, for instance, from cryptographically signed statements of bank accounts or insurance claims and their use in verifiable differentially private surveys could have considerable economic potential, as data markets require technologies that provide verifiability despite privacy protection~\cite{garrido2021revealing}.

There are two main approaches for verifiable computation: trusted execution environments (TEEs)~\cite{OMTPTEE}, and non-interactive or interactive \acp{ZKP}~\cite{goldwasser1989knowledge, simari_primer_nodate,ben2019scalable}. 
Given the numerous known vulnerabilities and attacks on TEEs~\cite{TEE_vulnerabilities, TEE_attacks_1, TEE_attacks_2} and Intel's SGX SDK deprecation~\cite{sgx}, we decided to focus on \ac{ZKP}-based approaches. Moreover, non-interactive ZKPs do not require to engage into sequential messaging, so -- unlike with interactive ZKP -- the prover can convince multiple parties of a claim with a single message~\cite{simari_primer_nodate}.
Thus, we opted to use non-interactive ZKPs to enable the verifiability of the computational integrity in the selected DP mechanism.

This paper's scope covers both binary answers, e.g., ``\textit{Are you older than $18$?}'', and numerical answers, e.g., ``\textit{How old are you?}''.
We provide plausible deniability for interviewees with \ac{DP} mechanisms in the local model, specifically, employing randomized response~\cite{warner1965randomized} and exponentially distributed noise~\cite{dwork2006noise}.
Lastly, we adapt these mechanism such that we can verify their correct execution with \acp{ZKP} by employing \acp{SNARK}~\cite{groth2006perfect, ben2013snarks}, resulting in the primitives represented in Algorithms~\ref{alg:verifiable_unif_random} and~\ref{alg:verifiable_exponential}.
We implement the corresponding circuits and evaluate their performance characteristics to assess our approach's practicality.\footnote{The source code can be found at \url{https://github.com/applied-crypto/DPfeatZKP}.} 

As randomized response and exponential noise are building blocks for other more complex mechanisms, our scheme could also be extended to prove their verifiability, such as in two-stage randomized response models~\cite{RR_two_stage_1, RR_two_stage_2}, unrelated question models~\cite{RR_question_model}, forced response models~\cite{RR_forced_response}, \ac{LDP} models~\cite{RR_local_DP_exterme_points, RR_local_DP_exterme_points_graphs, RR_DP_extremal, heavy_hitter_agg, binary_mech}, private weighted histogram aggregation in crowdsourcing by leveraging multivariate randomized response~\cite{yang_wireless_2016}, building histograms~\cite{DP_precision_Geometric}, or using exponential noise distributions in the central model of \ac{DP}. 
Such verifiable forms of \ac{DP} are also relevant in multilateral protocols that provide economic incentives for participation based on the participants' contribution.
In such settings, one should compute fair rewards from the original data without noise, requiring that the computation of both their deterministic contribution and the shared noisy value is verifiable. An example for such a scenario is fair blockchain-based federated learning, studied by Rückel et al.~\cite{rueckel2022fairness}.

\smallskip

This paper is structured as follows.
We provide preliminaries in Section~\ref{sec:Preliminaries}, discuss the SNARK-based approach and its implementation in Section~\ref{sec:Implementation}, and evaluate it in Section~\ref{sec:Evaluation}.
Lastly, we comment on related work in Section~\ref{sec:Related_work}, discuss our approach in Section~\ref{sec:Discussion}, and conclude the paper in Section~\ref{sec:Conclusion}.

%% file: 002_Preliminaries.tex
\section{Preliminaries}
\label{sec:Preliminaries}

\subsection{Differential Privacy}
\label{subsec:Differential_privacy}

We consider a collection of records from a population (dataset)~$D$ to belong to the universe of possible datasets~$\mathcal{D}$.
We let $D'\sim D$ denote neighboring datasets, i.e., $D$ and $D'$ differ by only one record.
Differential privacy, introduced by Dwork~et~al. in~2006~\cite{DP_original}, formalizes a mathematical definition of privacy whereby an analysis' output distribution is nearly the same across all neighboring datasets.
The indistinguishability between datasets is parameterized by $\varepsilon>0$. 
The higher $\varepsilon$, the easier it is to identify datasets. 

\begin{definition}
\label{def:DP}
  (($\varepsilon$, $\delta$)-Differential Privacy~\cite{dwork_algorithmic_2013}). A randomized mechanism $\mathcal{M}$ is ($\varepsilon$, $\delta$)-differentially private iff for any neighboring dataset $D'\sim D$, and any set of possible outputs $\mathcal{S}\subseteq
  Range(\mathcal{M})$,
  \begin{center}
  Pr[$\mathcal{M}$($D$)  $\in$  $\mathcal{S}$] $\leq$ $e^\varepsilon$ $\cdot$ Pr[$\mathcal{M}$($D'$)  $\in$  $\mathcal{S}$] $+$ $\delta$.      
  \end{center}
\end{definition}

Having a non-zero $\delta$ relaxes the strict $\varepsilon$ bound for possible but unlikely events; this type of guarantee is called \emph{approximate} \ac{DP}, whereas with $\delta = 0$, we obtain \emph{pure} \ac{DP}.
A randomized mechanism~$\mathcal{M}$ typically ensures \ac{DP} by adding carefully calibrated random noise to the output of a deterministic function $f(\cdot)$, for example, by adding exponentially distributed noise~\cite{dwork2006noise} to a count, average, or median.
Furthermore, another factor beyond $\varepsilon$ that calibrates noise is the sensitivity of $f(\cdot)$, which measures the maximum variation of the output as the input dataset $D$ changes (denoted as $\Delta$).

Lastly, it is important to note that a \ac{DP} mechanism $\mathcal{M}$ follows \textit{sequential composition} \cite{dwork_algorithmic_2013}, i.e., if $\mathcal{M}$ is computed $n$ times over a dataset $\mathcal{D}$ with $\varepsilon_i$, in effect, the total $\varepsilon$ is given by $\sum \varepsilon_i$.
Thus, the results become less private with every query.
Yet, a system can effectively impede an attacker from averaging out the noise through a sequence of DP results by blocking subsequent queries or deterministically generating the randomness based on the query parameters.

\subsection{Local Differential Privacy}
\label{subsec:Local_Differential_privacy}

Definition~\ref{def:DP} corresponds to the \emph{central model}, in which a trusted curator collects data points and adds noise from a distribution whose variance is tuned by the function's sensitivity ($\Delta$) and the required degree of plausible deniability ($\varepsilon$). 
In this paper, we focus on the \emph{local model}, whereby the data subject obfuscates the data points directly before sharing. While the local model typically provides less accuracy than the central model, the data subject does not need to trust the curator from a privacy perspective.

For the local setting, we adopt similar notation to~\cite{geometric_mechanism,kacem_geometric_2018}.
We let $\mathcal{X}$ contain all the possible records of a population, where $x\in \mathcal{X}$ holds a particular individual's data.
We let $f:\mathcal{X}\to [l,u]$ be a function that maps each element of $x \in \mathcal{X}$ to $f(x) \in [l,u]$, where $[l,u]$ is the set of integers between the lower bound~$l\in\mathbb{N}$ and the upper bound~$u\in\mathbb{N}$.
In practice, $f(\cdot)$ provides information about an individual, e.g., their age, income, height, or blood pressure.
Let $\mathcal{M}:f(x) \to [0,n]$ denote a randomized mechanism that maps each deterministic query output $f(x) = i\in [l,u]$ to each possible value $j\in[l,u]$ following a probability distribution that depends on the value $i$. 
In this setting, a randomized mechanism $\mathcal{M}$ provides local ($\varepsilon$, $\delta$)-DP iff for every pair of inputs $x, x' \in \mathcal{X}$, and for every possible output $j\in[l,u]$: 
\begin{center}
    Pr[$\mathcal{M}$($x, f$)  $=$  $j$] $\leq$ $e^\varepsilon$ $\cdot$ Pr[$\mathcal{M}$($x', f$)  $=$  $j$] $+$ $\delta$.  
\end{center}
 
We will cover two mechanisms that satisfy local DP: randomized response~\cite{warner1965randomized} for binary data and exponentially distributed noise~\cite{dwork2006noise} for numerical data. 
\smallskip

\noindent \textbf{Randomized response.} Warner~\cite{warner1965randomized} introduced randomized response in 1965 to provide plausible deniability to interviewees, which encouraged them to answer truthfully, thus reducing bias in surveys. 
The interviewees would answer queries $f(\cdot)$ of the form ``\textit{Are you a} [...]?". The following algorithm is well-known to be ($\ln3$,~$0$)-differentially private~\cite{dwork_algorithmic_2013}:
\begin{enumerate}
    \item Flip a coin. 
    \item If heads, answer truthfully.
    \item Else, flip the coin again and answer ``Yes" if heads and ``No" otherwise.
\end{enumerate}


\noindent \textbf{Exponentially distributed noise.}
Early work by Dwork et al.~\cite{dwork2006noise} shows that noise distributed as $\mathrm{Pr}[x] \propto \exp(-\tfrac{\varepsilon|x|}{\Delta})$ fulfills ($\varepsilon$, $0$)-DP.
We leverage exponentially distributed noise for locally obfuscating numerical data to answer queries $f(\cdot)$ of the form ``\textit{How many} [...]?'', i.e., count queries, by adding noise to the deterministic output of $f(\cdot)$ in this manner: $\mathcal{M}(x, f)= f(x) + \mathrm{\textbf{noise}}$. 
As the responses are local, we must ensure indistinguishability between any $f(x)$ and $f(x')$; thus, we must set $\Delta$ according to the output range of $f(\cdot)$.
In practice, to ensure that an attacker cannot easily distinguish individuals in the extreme-case scenario, e.g., a newborn from a $128$-year old person, with the query ``\textit{How old are you}?,'' we set $\Delta = |\max\limits_{x} f(x) - \min\limits_{x} f(x)| = |128 - 0|$.

As presented above, both mechanisms fulfill pure DP; however, generating exponentially distributed noise is subject to approximation errors in practical implementations and, thus, we only achieve approximate \ac{DP} instead~\cite{gazeau2016preserving}. 
We shed more light on this issue in Section~\ref{sec:Implementation}. 

\subsection{Proof Systems and SNARKs}
\label{subsec:zkps}

There have been several key milestones in the work towards cryptographically verifiable computations.
Babai~\citep{babai1985trading} studied interactive proofs between a \textit{prover} and a \textit{verifier} and analyzed which problems can be checked by a polynomially bounded verifier when adding randomness and interaction. Fiat and Shamir~\citep{fiat1986prove} then introduced a heuristic to replace the verifier with a random oracle, which one can implement with a secure hash function. Nonetheless, one uses the word ``argument'' instead of ``proof'' in this case because the existence of a secure hash function has not been proven mathematically so far and is rather a working hypothesis, also bound to today's compute and (differential) cryptanalysis capabilities.
\citet{goldwasser1989knowledge} proved that one could verify a large class of problems probabilistically in this way, where the verifier additionally does not even need to learn anything beyond the statement's correctness. While practical applications of the accordingly termed \acp{ZKP} were rare after these early developments, a period of rapid improvements started in the mid-2000s and led to the computationally efficient (quasi-linear complexity) generation of succinct arguments of logarithmic size and verification time, called \acp{SNARK}~\cite{groth2006perfect, ben2013snarks}. 

The research community has since developed other flavors such as \acp{STARK}, which differ in the setup procedure, proof size, and cryptographic assumptions, but have similar functional aspects.
These new approaches allow succinct or scalable zero-knowledge proofs for the correctness of arbitrary statements and, thus, practical verifiable computation, i.e., proofs for the correct execution of a program without displaying all inputs, outputs, or intermediate steps. Additionally, one can reveal Merkle proofs or other cryptographic relations like the public keys corresponding to a digital signatures by a reputed institution on the inputs to force the prover to use unknown but fixed variables. 
Arguably, the core area of application of \acp{ZKP} today is in distributed ledgers, where because of redundant execution, cheap (succinct) verification without revealing sensitive data is important to solve scalability issues and mitigate excessive data visibility~\citep{ben2019scalable,sedlmeir2022transparency}. 

%% file: 003_Implementation.tex
\section{Implementation}
\label{sec:Implementation}

In this section, we first describe how to adapt standard implementations of uniform randomness generation, randomized response, and exponentially distributed noise (see Section~\ref{sec:Preliminaries}) such that \acp{ZKP} can verify their use. 
For our implementation, we employ Circom, a well-known, open-source technology stack for implementing \acp{ZKP}~\cite{circom}.
Circom is a domain specific programming language and compiler that translates JavaScript-like arithmetic circuits in a \ac{R1CS}, on behalf of which further libraries (e.g., SnarkJS) can generate \acp{SNARK}.
A circom-specific variable type to explicitly define constraints is called \emph{Signal}. The programming of these signals is restricted by the underlying \ac{QAP}, which the \ac{R1CS} encodes, to use only quadratic constraints inside one Signal. Therefore, a Signal can only be assigned once and is immutable. For this reason, calculations often have to be split into multiple sub-calculations. Moreover, branchings and loops can only be used in a restricted way, for instance, the maximum number of iterations must be specified by a constant instead of a variable or Signal. 

\begin{table}[!t]
\centering
\caption{Notation}
\begin{tabular}{c  l} 
\hline
\vspace{1.5pt}
 $\mathcal{X}$ & Universe of records \\ \vspace{1pt} 
 $x  \in \mathcal{X}$  & Individual record \\ \vspace{1pt} 
 $f: \mathcal{X} \to [l, u] $  & Query function \\ \vspace{1pt}
 $l$ & Lower bound, $\min\limits_{x} f(x)$  \\ \vspace{1pt}
 $u$ & Upper bound, $\max\limits_{x} f(x)$  \\ \vspace{1pt}
 $\mathcal{M}:[l,u] \to [l,u]$ & Randomized mechanism \\ \vspace{1pt}
 $\Delta$ & Sensitivity of $f(\cdot)$, $|u-l|$  \\ \vspace{1pt}
 $\ell$ & Noise added to $f(\cdot)$ \\ \vspace{1pt}
 \verb+nBits+ & Number of bits representing $\ell$  \\ \vspace{1pt}
 $p_k$ & Bias of bit $k$, $\mathrm{Pr}[\ell_k=1]$  \\ \vspace{1pt}
 $d$ & Precision \\ 
\hline 
\end{tabular}
\label{table:Notation}
\end{table}

In what follows, we will use the roles prover for the survey participant and verifier for the surveyor, and the notation specified in Table~\ref{table:Notation}. We also assume that both of them have a dedicated key-pair that they can use for end-to-end encrypted, authenticated communication and for recognizing each other, or another means to bind them to a specific secret key, for instance, through an anonymous credential with private holder binding~\citep{camenisch2001efficient}. Such anonymous credentials can be implemented with specific-purpose \acp{ZKP}~\cite{camenisch2001efficient} and with \acp{SNARK}~\cite{delignat2016cinderella,schanzenbach2019zklaims}. Key-pairs are a common way to facilitate the generation of verifiable randomness, for instance, in Algorand's consensus mechanism~\cite{chen2019algorand}.

\subsection{Verifiable Uniform Randomness}
\label{subsec:UnifDistRand}

To achieve uniform randomness that cannot be spoiled unilaterally by either the prover or the verifier, we employ two inputs and a hash function as a random oracle~\citep{canetti2004random}. 
More specifically, we sign a challenge that the verifier specifies with the prover's private key and hash the result. As the private key is determined by the fixed prover's public key, neither of the two parties can bias the resulting randomness without collusion. 
We use Poseidon\footnote{Using other hashing mechanisms is possible, yet the performance can become considerably worse -- for instance, in the case of SHA256, around~30x.} -- a relatively new hashing algorithm that was specifically developed for use in \acp{ZKP} and that is already being used in many blockchain-based applications on Ethereum and, therefore, to some extent battle-tested~\cite{grassi2021poseidon}. 
We represent this building block as a function in Algorithm~\ref{alg:verifiable_unif_random} between lines $1$ and $4$, using existing components in Circom for EdDSA signature verification, Poseidon, and conversion of (large) integers to binary representation. 
Assuming that the Poseidon hash function is a random oracle and the keypair was created without anticipating the survey and the verifier's challenge, this gives us an array of $254$~unbiased random bits.

\SetKwComment{Comment}{// }{}
\RestyleAlgo{ruled}
\begin{algorithm}[t!]
\footnotesize
\DontPrintSemicolon
\caption{Verifiable randomized response for binary data and uniform randomness (``unbiased coins'').}
\label{alg:verifiable_unif_random}

\KwData{$v$: binary truthful value (``Yes'' or ``No''); $a$: prover contribution to randomness (secret key); $b$: verifier contribution to randomness (challenge).}
\KwResult{Differentially private answer.}

\SetKwFunction{Fsub}{VerifiableUnifRand}
\SetKwProg{Fn}{Function}{:}{}
\Fn{\Fsub{$a$, $b$}}{
    $s$ = sign($a$, $b$)\Comment*[r]{sign challenge with secret key}
    $r$ = hash($s$) \Comment*[r]{$r$ is an array of bits}
    \KwRet $r$\;
}

\SetKwFunction{FMain}{VerifiableRandomizedResponse}
\SetKwProg{Fn}{Function}{:}{}
\Fn{\FMain{$v$, $a$, $b$}}{
$r$ = \Fsub{$a$, $b$}\;
\uIf{$r[0] = 0$}{ 
  \KwRet $v$\;
  }{\uElseIf{$r[1] = 0$}
    {\KwRet No\;}
    \Else{\KwRet Yes\;} 
  }
  }

\end{algorithm}

\subsection{Verifiable Randomized Response}
\label{subsec:Randomized_response}

Randomized response is simple to verify with \acp{ZKP} by utilizing the verifiable uniform randomness function (see Algorithm~\ref{alg:verifiable_unif_random}).
In practice, without loss of generality, we only consider the least two significant bits of the random number generated.
For the randomized response algorithm presented in Section~\ref{sec:Preliminaries} and presented formally in Algorithm~\ref{alg:verifiable_unif_random}, we need to sample at least once (last bit) and at most twice (second-last bit), depending on the first coin flip. 
The source code from Fig.~\ref{code:randomizedResponseCircom} implements this in Circom. As if-statements are not natively possible in \acs{R1CS} and, therefore, only available with restrictions in Circom, we arithmetize the corresponding statements in lines $7$ to $12$ from Algorithm~\ref{alg:verifiable_unif_random} into the lines $39$ to $40$ from Fig.~\ref{code:randomizedResponseCircom}. 

\begin{figure}[!htb]
\begin{lstlisting}[language=circom, basicstyle=\scriptsize\ttfamily, escapeinside={(*}{*)}]
pragma circom 2.0.0;

include "./poseidon.circom";       // Poseidon hashing
include "./bitify.circom";         // Bit array conversion
include "./eddsaposeidon.circom";  // Signature checking

template Main() {
   signal input value;        // v
   signal input challenge;         
   signal input R8[2];        // elliptic curve element of signature
   signal input S;            // field element of signature
   signal input pk[2];        // public key

   // check signature on challenge against public key
   component eddsaVerifier = EdDSAPoseidonVerifier();
   eddsaVerifier.Ax <== pk[0];
   eddsaVerifier.Ay <== pk[1];
   eddsaVerifier.S <== S;
   eddsaVerifier.R8x <== R8[0];
   eddsaVerifier.R8y <== R8[1];
   eddsaVerifier.M <== challenge;
   eddsaVerifier.enabled <== 1;  // checks signature implicitly

   // hash signature and convert this randomness to bit array
   component hash = Poseidon(3);
   component bitify = Num2Bits_strict();
   hash.inputs[0] <== R8[0];
   hash.inputs[1] <== R8[1];
   hash.inputs[2] <== S;
   bitify.in <== hash.out; 
   signal randSeq[254];
   for(var i = 0; i < 254; i++) {
      randSeq[i] <== bitify.out[i];
   } 

   // determine result from randomness
   signal rand; 
   signal output out;
   rand <== randSeq[0] * randSeq[1];
   out <== (1 - randSeq[0]) * value + rand;
   
}

component main {public [challenge, pk]} = Main();

\end{lstlisting}
\caption{Circom code for a component that implements verifiable randomized response.}
\label{code:randomizedResponseCircom}
\end{figure}

\subsection{Verifiable Exponentially Distributed Noise}
\label{subsec:Geometric_mechanism}

The exponentially distributed noise adaptation to \acp{ZKP} is not as straightforward as with randomized response because it typically involves floating point operations and rounding. 
After trying different implementations of exponentially distributed noise generation -- we briefly cover the journey in Section~\ref{sec:Discussion} --we successfully adapted the method proposed by Dwork et al.~\cite{dwork2006noise} to \ac{ZKP}, which we present in Algorithm~\ref{alg:verifiable_exponential}:
In their method, Dwork et al. approximated exponentially distributed noise of $\mathrm{Pr}[x] \propto \exp(-\tfrac{\varepsilon|x|}{\Delta})$ with the Poisson distribution, fulfilling ($\varepsilon$, $\delta$)-DP.
Their method samples noise by producing a sequence of biased bits equal in number to the number of bits in the binary expansion of the noise $\ell$. The algorithm flips an extra bit to add a sign ($\pm \ell$). The bias of each bit $k \in \{0, ... ,$ \verb+nBits+$\}$ representing $\ell$ in binary is given in Section 4.1 of \cite{dwork2006noise} by 
$$
\mathrm{Pr}[\ell_k]:=\mathrm{Pr}[\ell_k=1] = \left(1 + \exp\left(\tfrac{\varepsilon\cdot 2^k}{\Delta}\right)\right)^{-1}.
$$

To generate biased bits from unbiased bits, we include in Algorithm~\ref{alg:verifiable_exponential} a well-known technique: first, we expand in binary the bias $p_k$ of a bit $k$. 
Afterward, the algorithm sequentially examines random unbiased bits until one differs from the corresponding bit in the binary expansion of $p_k$ and, subsequently, outputs the complement of the random unbiased bit~\cite{dwork2006noise}. 
Essentially, this approach allows to simulate biased coins up to a pre-defined precision with unbiased coins. 
However, the method employed has three limitations.

The \emph{first} limitation entails several issues that relate to representing with a limited precision $d$ the bias of the bits composing $p_k$, i.e., $d$ is the number of bits available for representation.
Nonetheless, the probability of the inner loop not terminating for $j<d$ and, therefore, raising an error decays with $2^{-d}$. Thus, we can easily choose $d$ such that the likelihood of this event is negligible (line $17$ of Algorithm~\ref{alg:verifiable_exponential}).
Furthermore, we show that we can provide enough precision in our circuit:
The randomness generated from a single Poseidon hash could provide a precision of around $2^{252}\approx10^{75}$, i.e., $d\approx75$. 
By using multiple rounds of hashing and signing, we could also generate more random bits and account for higher precision needs.
Additionally, we restrict noise values $\ell$ to the interval $[l, u]$, where $u$ and $l$ are the deterministic function's output upper and lower bounds, respectively.
For our experiment on polling individuals' age, we employed the algorithm with $d=20$ and $\Delta=|u-l|=128$.
These example values require the algorithm to represent $\ell$ with \verb+nBits+=$7$ bits and, in turn, generate one instance of noise with $d\cdot 7+1=141<256$ bits (``times $d$'' because each of the 7 bits' bias will be expanded to $d$ bits and one more for the sign). 
In other words, a single round of hashing and signature verification is sufficient (and would still be sufficient for $d=35$, which corresponds to an error bound of $2^{-35}\approx 10^{-10}$ when approximating probabilities~\cite{dwork2006noise}), and a negligible probability to raise an error (upper bound $7\cdot 2^{-20}$). We could also aim for the typical machine accuracy of $10^{-16}$ by using $d>16\cdot \log_2(10)\approx 53.1$, i.e., $d\geq54$, which would involve the creation of two independent random bit arrays.

These design decisions allow us to approximate the Poisson distribution with an error bound that we can determine and control ex-ante, when designing the survey. 
Thus, we achieve ($\varepsilon$, $\delta$)-DP with a statistical difference of $\delta = $ \verb+nBits+$\cdot 2^{-d} = 7\cdot2^{-20}$~\cite{dwork2006noise}. 
Consequently, for improving the DP guarantee on $\varepsilon$, we only need to increase $d$.
Moreover, the probability mass outside the considered interval $[-2^d$, $2^d]$ is redistributed inside the interval, leading to an additional statistical difference of $2\exp(-(\varepsilon\cdot 2^d)/\Delta)$ that we let the term \verb+nBits+ absorb~\cite{dwork2006noise}.

The \emph{second} limitation is the zero probability assigned to noise values of a binary expansions with more bits than \verb+nBits+ (i.e., noise outside of $[l, u]$).
Dwork et al. proposed to constrain the algorithm's output, i.e., deterministic answer + \textbf{noise}, to \verb+nBits+ and return the deterministic answer in case there is an overflow. 
According to Dwork~et~al.~\cite{dwork2006noise}, as the distribution in the range $[l, u]$ is exponential, we maintain the same privacy guarantee by increasing the probability of not adding noise by a \emph{trivial} amount (i.e., $\delta$ increases). 
We execute a modulo operation to remap any output value outside $[l, u]$ back in that range to reduce such an increase (lines $21$ and $23$ of Algorithm~\ref{alg:verifiable_exponential}).
Intuitively, a modulo operation on the output preserves DP as it is a post-processing step and, also, will re-distribute the outputs in the range instead of on one value.
Formally, the proof may be found in Lemma~$3$ of Wang~et~al.~\cite{LunWangUCB}.

\begin{figure}
    \centering
    \includegraphics[width=0.9\linewidth]{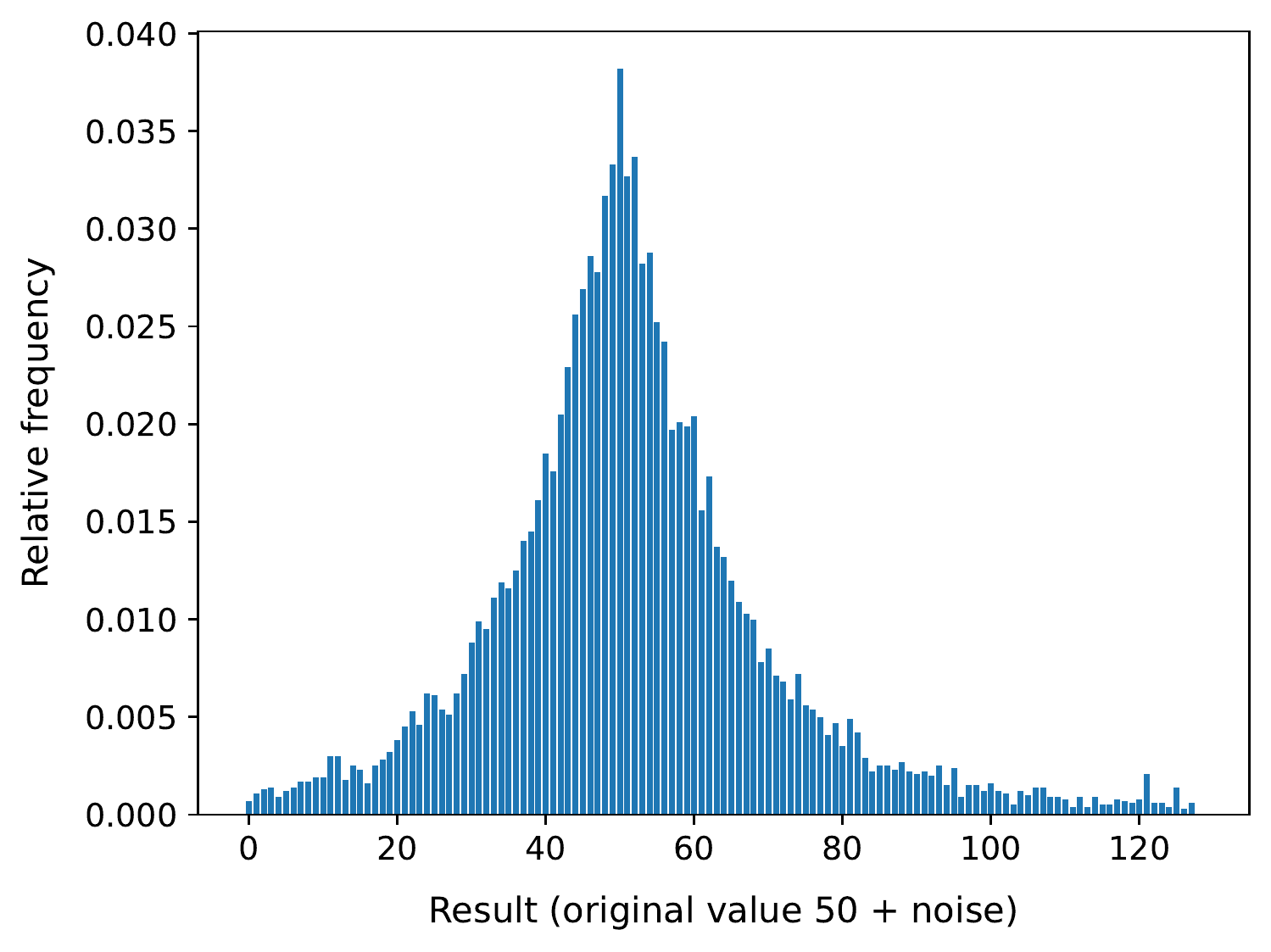}
    \caption{Example histogram for $\mathbf{l=0}$, $\mathbf{u=128}$, $\mathbf{d=20}$, $\boldsymbol{\varepsilon}\mathbf{~=10}$, and true value $\mathbf{v=50}$ with a sample size of $\mathbf{10\,000}$.} 
    \label{fig:histogram}
\end{figure}

The \emph{third} limitation comes with flipping an unbiased bit to assign the sign of the noise, which converts a Poisson distribution into a two-sided distribution with double the probability on its center, i.e., of noise~$=0$.
While Dwork~et~al.~did not address this issue in~\cite{dwork2006noise}, we could follow the approach of Champion~et~al.~\cite{reject_sample_coin_DP} of rejecting $-0$ and executing the algorithm again (section~$3.3$ of~\cite{reject_sample_coin_DP}).
DP is maintained as the number of failures is independent of the noise. 
However, instead, to remain computationally performant, we output a uniformly sampled value within $[l, u]$ if $-0$ (lines $19$ to $23$ of Algorithm~\ref{alg:verifiable_exponential}), effectively removing the excess probability at $0$.
Intuitively, we preserve DP by adding more noise to the output distribution.
Formally, we provide this justification: Let $P_{\mathrm{old}}$ be some DP distribution on $N$ discrete values and $P_{\mathrm{new}}$ such that $P_{\mathrm{new}}(x) = \alpha P_{\mathrm{old}}(x) + (1-\alpha)/N$.
Obviously, this is a probability distribution. 
In our case, $1-\alpha$ is the probability of obtaining noise $-0$.
For any $D$ and $D^\prime$ that differ in at most one record,
\begin{align*}
&\mathrm{Pr}[\mathcal{M}_{\mathrm{new}}(D)\in \mathcal{S}]\\
&= \alpha\cdot \mathrm{Pr}[\mathcal{M}_{\mathrm{old}}(D)\in \mathcal{S}] + (1-\alpha)\cdot\frac{|D|}{N} \\
&\leq\alpha e^{\varepsilon}\cdot \mathrm{Pr}[\mathcal{M}_{\mathrm{old}}(D^\prime)\in \mathcal{S}]
+ \alpha\delta + (1-\alpha)\cdot\frac{|D|}{N}\\
&=\alpha e^{\varepsilon}\cdot\left(\frac{\mathrm{Pr}[\mathcal{M}_{\mathrm{new}}(D^\prime)\in\mathcal{S}]}{\alpha} - \frac{|D^\prime|\cdot(1-\alpha)}{N\alpha}\right) 
+ \alpha\delta + (1-\alpha)\cdot \frac{|D|}{N}\\
&= e^\varepsilon\cdot \mathrm{Pr}[\mathcal{M}_{\mathrm{new}}(D^\prime)\in\mathcal{S}] + \alpha\delta + (1-\alpha)\cdot\frac{|D|-e^\varepsilon|D^\prime|}{N}\\
&\leq e^\varepsilon\cdot \mathrm{Pr}[\mathcal{M}_{\mathrm{new}}(D^\prime)\in\mathcal{S}] + \alpha\delta + (1-\alpha)\cdot\frac{1}{N}.
\end{align*}
Given that $\varepsilon\geq0$, then $e^\varepsilon\geq1$. Moreover, $|D|-|D^\prime|\leq1$ because they are neighboring. Thus, since we choose to re-distribute the excess weight for noise~$-0$ ($\alpha = 1 - \frac12 \mathrm{Prob}(0)$), $\delta$ may grow to at most $\frac12\mathrm{Prob}(0)\cdot\frac{1}{N}$.
The above formulation is a universal upper bound: Essentially, it proves that the convex combination of an $(\varepsilon,\delta_1)$ mechanism and a uniform distribution with pointwise weight $\delta_2=\frac1N$ (which is obviously $(0,\delta_2)$-DP) is $(\varepsilon,\delta)$-DP, where $\delta$ is the convex combination of $\delta_1$ and $\delta_2$. Future work could focus on obtaining a tighter $\delta$ bound specifically for the Poisson distribution or employ the method from Champion~et~al.~\cite{reject_sample_coin_DP}, which would maintain a $\delta$ upper bound of $\verb+nBits+\cdot 2^{-d}$.
Altogether, this approximation allows us to sample exponentially distributed noise preserving ($\varepsilon$, $\delta$)-DP in a way that we can successfully verify with \acp{ZKP}. 
We depict an example of the resulting output distribution in Fig.~\ref{fig:histogram}.

As Circom does not allow for branching, i.e., implementing conditional checks and breaking or continuing loops, besides the workaround for if-statements, we had to introduce some additional Signals (see Fig.~\ref{code:exponentialCircom}). These Signals allow us to determine the correct return value although all iterations from the loop are simulated in the circuit.
We did this by introducing Signal arrays $\verb+hit+ _{1 \ldots \text{nBits}}$ with length $d$ that help to identify the firsts unequal pair of bits in one~($j$) loop, where the loop would break in Algorithm~\ref{alg:verifiable_exponential}. 
This approach ensures that only the first occurrence of unequal bits (index $i$) is taken into account for the calculation of the biased randomness.
Moreover, we multiply the inverted binary result from \verb+isEqual+, which compares the corresponding bits of the random sequence $r_j$ and the probability $B_{j, p_k}$, with the bit of the probability and the hit bit-value, which is $1$ as long as the result of \verb+isEqual+ of the last iteration of the loop was $1$, essentially $[1_{0}, \ldots, 1_{i-1}, 1_{i}, 0_{i+1}, \ldots, 0_{d-1}]$.
Therefore, only at the first inequality of those two bits the probability bit is not multiplied with zero, and, thus, can be taken into account for the noise. In other words, $eval3[k][j]$ is the ``running return value'' after the $j+1$st iteration of the loop, and is set to $1$ only if the $j$th bit of the probability is one, the $j$th bit of the random sequence is $0$, and the first time that the probability and random bit array are different occurs at position $j$ as well.

\SetKwComment{Comment}{// }{}
\RestyleAlgo{ruled}
\begin{algorithm}[tb!]
\footnotesize
\DontPrintSemicolon
\caption{Verifiable exponentially distributed noise generation for numerical data. By $x~\mathrm{mod} (l,u)$ we denote $l + \left(x~\mathrm{mod} (u-l)\right)$.}
\label{alg:verifiable_exponential}

\KwData{$v$: integer-valued truthful value; $u$: upper bound; $l$, lower bound; $\Delta = |u - l| \geq 0$: sensitivity of query function; $\varepsilon \geq 0$: privacy parameter; $d \geq 0$: precision of binary expansion; $a$: prover contribution to randomness (secret key); $b$: verifier contribution to randomness (challenge).}
\KwResult{$v+\mathrm{noise}\sim\mathrm{Pois}(v|\tfrac{\varepsilon}{\Delta})$.}

\SetKwFunction{FMain}{VerifiableExponentialNoise}
\SetKwProg{Fn}{Function}{:}{}
\Fn{\FMain{$v$, $\Delta$, $\varepsilon$, $d$}}{
$B_K = \mathrm{BinaryExpansion}(\Delta)$\;
$B_v = \mathrm{BinaryExpansion}(v)$\;
$B_r = []$ \Comment*[r]{$B_r$ stacks biased bits}
\For{$k\leftarrow 0$ \KwTo $\mathrm{NumBits}(B_K)$}{
    $p_k = \tfrac{1}{1+\exp(2^k\tfrac{\varepsilon}{\Delta})}$\;
    $B_{p_k} = \mathrm{BinaryExpansion}(p_k)$\;
    \Comment{$r$ has at least $d$ bits}
    $r =$ \Fsub{$a$, $b$}\; 
  \For{$j\leftarrow 0$ \KwTo $d$}{ \Comment {Where $d$ is the least significant bit}
    $r_j =$ r[j] \Comment*[r]{$r_j \in \{0, 1\}$}
    \eIf{$r_j = B_{j,p_k}$}{ 
      \textbf{continue}\;}{
      $B_r.\mathrm{push}(B_{j,p_k})$\; 
      \textbf{break}\;
    }
    \If{$j = d$}
    {\KwRet $\mathrm{RaiseError}$}
    }
}
{$\mathrm{noise} = \mathrm{DecimalExpansion}(B_r)$\;}
{$\mathrm{sign} = \mathrm{VerifiableUnifRand}(a, b)[0]$\;} 
\eIf{$(\mathrm{noise} = 0 \mathrm{~and~} \mathrm{sign} = 0)$}{
  \KwRet $\mathrm{DecimalExpansion(VerifiableUnifRand}(a, b)) \mathrm{~mod~} (l,u)$\;}{
  \KwRet  $\left[(v + (2\cdot\mathrm{sign} - 1)\cdot\mathrm{noise}\right] \mathrm{~mod~} (l,u)$\; 
}
}
\end{algorithm}

\begin{figure}[!htb]
\begin{lstlisting}[language=circom, basicstyle=\scriptsize\ttfamily, escapeinside={(*}{*)}]

// include statements as before, plus modulo component 
template Main(nBits, d) {
   signal input challenge, value
   signal input prob[nBits][d];  // binary expansions of p_k
   signal input R8[2], S, pk[2]; // signature and public key
   // check the EdDSA signature of the challenge against pk and put it in the hash component to create randSeq, as in Figure 1 (lines 10 to 34). 
   ...

   component isEqual[nBits][d];
   signal noiseBits[nBits];
   signal eval1[nBits][d];
   signal eval2[nBits][d];
   signal eval3[nBits][d + 1];
   signal hit[nBits][d + 1];

   // run the algorithm to create biased coins
   for (var i = 0; i < nBits; i++) {
      for (var j = 0; j < d; j++) {
         isEqual[i][j] = IsEqual();
      }
   }
   for (var k = 0;  k < nBits; k++) {
      hit[k][0] <== 1;
      eval3[k][0] <== 0;
      for (var j = 0;  j < d; j++) {
         isEqual[k][j].in[0] <== prob[k][j];
         isEqual[k][j].in[1] <== randSeq[k * d + j];
         hit[k][j + 1] <== 
         hit[k][j] * isEqual[k][j].out;
         eval1[k][j] <== hit[k][j] * (1 - isEqual[k][j].out);
         eval2[k][j] <== eval1[k][j] * prob[k][j];
         eval3[k][j + 1] <== eval3[k][j] + eval2[k][j];
      }
      noiseBits[k] <== eval3[k][d];
   }

   component numify[2];
   // compute exponential noise from its binary representation 
   numify[0] = Bits2Num(nBits);
   for (var i = 0; i < nBits; i++) {
      numify[0].in[i] <== noiseBits[i];
   }
   signal absNoise <== numify[0].out; 
   signal positiveNoise <== randSeq[nBits * (d + 3)] * (value + absNoise);
   signal noisedResult <== (1 - randSeq[nBits * (d + 3)]) * (value - absNoise) + positiveNoise;   
   // generate uniformly distributed noise 
   numify[1] = Bits2Num(nBits);
   for (var i = 0; i < nBits; i++) {
      numify[1].in[i] <== randSeq[((d + 2) * nBits) + i];
   }
   
   component isZero = IsZero();  // check if noise == -0
   isZero.in <== absNoise;
   signal isUnif <== isZero.out * (1 - randSeq[nBits * (d + 3)]);
   signal unif <== isUnif * numify[1].out;
   signal result <== (1 - isUnif) * noisedResult + unif;

   component modulo = Modulo();
   modulo.in <== result;
   modulo.mod <== 128;
   signal output out <== modulo.out;
}

component main {public [challenge, pk]} = Main(7, 22);

\end{lstlisting}
\caption{Circom code for generating verifiable \ac{LDP} noise with $l=0$ and $u=128$. Import statements, signal definitions, and EdDSA verification omitted (see also Figure~\ref{code:randomizedResponseCircom}).}
\label{code:exponentialCircom}
\end{figure}

\subsection{Application: Verifiable Differentially-Private Polling with Anonymous Credentials}
\label{subsec:Anoncreds}

With the primitives presented in Algorithms~\ref{alg:verifiable_unif_random} and~\ref{alg:verifiable_exponential}, and an implementation of anonymous credentials with Circom, we can now implement verifiable, differentially private polling. 
Note that a Circom-based implementation of anonymous credentials allows to selectively disclose attributes from a digital certificate that corresponds to a Merkle tree with a signed root, and incorporate authenticity checks, private holder binding, expiration, revocation, and predicate proofs such as range proofs. 
We first sketch an example of a hypothetical setting. 
Digitally signed attestations of a person's attributes are stored in their ``digital wallet'' -- a mobile application -- in the form of an anonymous credential, which could contain personal information such as the holder's name, age, and gender, as well as a digital signature from an institution that the surveyor trusts regarding the authenticity of the information, e.g., a government or hospital.
The digital wallet can respond to so-called proof requests~\cite{schlatt2021kyc} that include requirements from the verifier's side what the survey participant should prove. 
In our case, this could include the following requirements:
\begin{itemize}
    \item Prove knowledge of (i) an authentic anonymous credential, issued by some institution, and (ii) knowledge of the secret key associated with the public key for the private holder, which is a binding included as one of the attributes in the anonymous credential.
    \item Prove that the anonymous credential is (i) not expired (range proof on expiration attribute) and (ii) not revoked (proof about set-inclusion or exclusion, referring to some public accumulator value as specified by the verifier).
    \item Reveal the result of our implementation of verifiable randomized response or exponentially distributed noise, applied to one of the (boolean or integer-valued) attributes in the credential. The attribute is represented by the issuer's signature on the anonymous credential, for instance, the attribute could be a leaf in a Merkle tree whose root is signed by the issuer.
    \end{itemize}
    
In the case of a \ac{SNARK}, the wallet (or the proof request) would also need to contain the structured reference string (proving key) generated in a setup procedure. 
It is important that while generally this proving key must be generated in a multi-party computation, in this case, it can be generated by the surveyor alone: Any party that knows how the structured reference string was created can fake proofs, but the privacy guarantees are not harmed in this case~\cite{fuchsbauer2018subversion}.

\begin{figure*}
    \centering
    \includegraphics[
    scale=0.19]{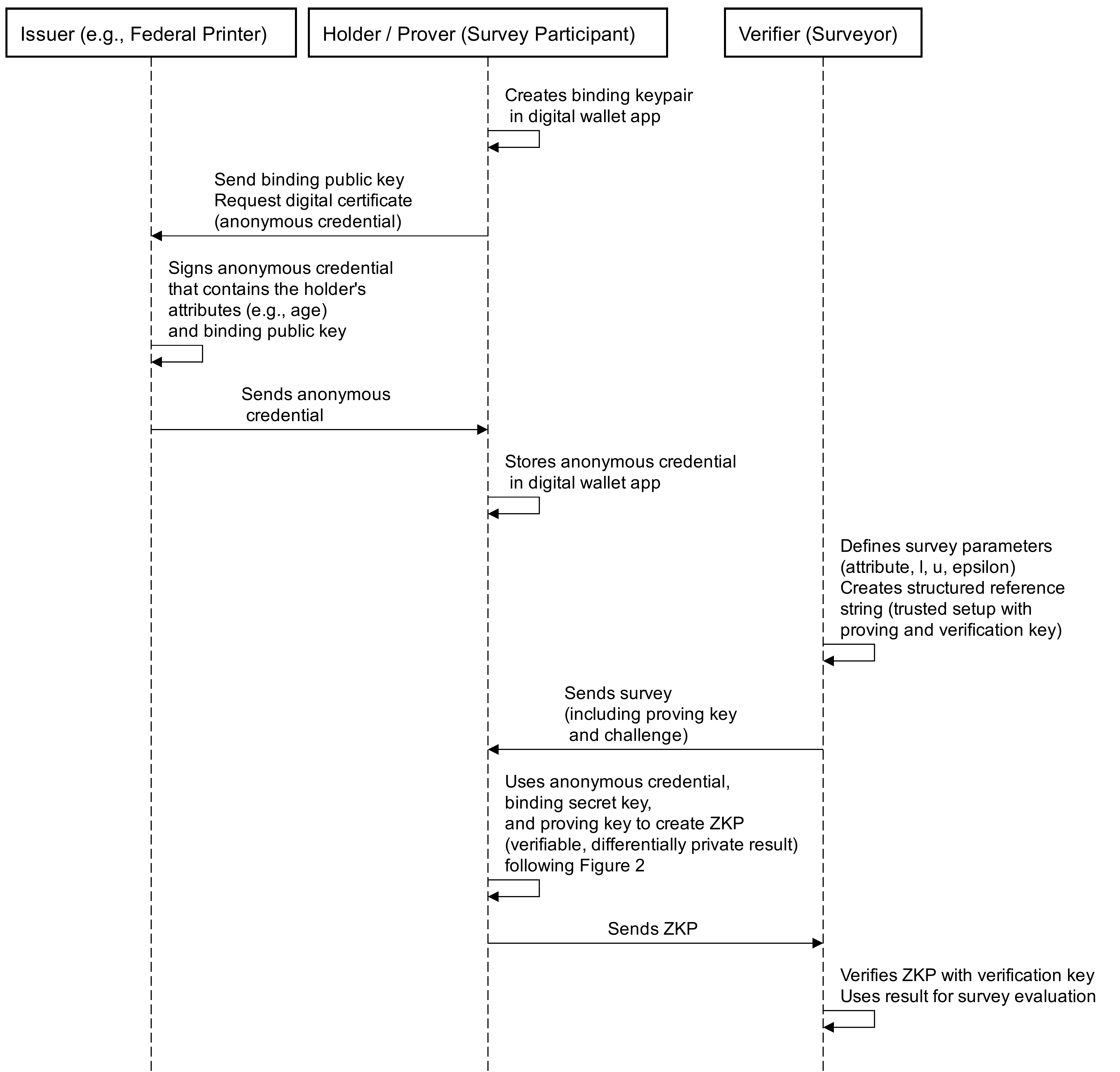}
    \caption{Process of participating in a survey with verifiable differential privacy.}
    \label{fig:workflow}
\end{figure*}

When the surveyor does not leak the ``toxic waste'' used for creating the structured reference string, the \ac{ZKP}'s soundness guarantee provides a chain of trust for the attribute, which is not directly revealed but modified through verifiable noise.
Lastly, the verifier (surveyor) can cryptographically check that the attribute and the survey participant's secret key for private holder binding are used as private inputs for the \ac{LDP} mechanism, and that the challenge as specified by the surveyor is used. We illustrate the survey process with anonymous credentials and verifiable differential privacy in Fig.~\ref{fig:workflow}.

%% file: 004_Evaluation.tex
\section{Evaluation}
\label{sec:Evaluation}
In this section, we discuss the performance and practicality of our approach for verifiable \ac{LDP}. 
We restrict the discussion to verifiable exponentially distributed noise because it includes strictly more complex operations, so by demonstrating its reasonable performance, we can conclude that verifiable randomized response also is practical.  
The implementation process of our verifiable \ac{LDP} approach with exponentially distributed noise was two-tiered. 
First, we implemented Algorithm~\ref{alg:verifiable_exponential} in Javascript and verified that it indeed yields an exponential distribution, where values that -- after adding the \ac{LDP} noise -- exceed the range of the output are instead displayed without added noise. 
Secondly, we employed the Poseidon hash function to create a random Oracle that generates verifiable randomness jointly from the prover's and the verifier's input. Next, we implemented the corresponding circuits in Circom.

Our choice of Algorithm~\ref{alg:verifiable_exponential} and our choice of implementation as displayed in Fig.~\ref{code:exponentialCircom} yielded a highly efficient implementation for creating verifiable, Poisson-distributed \ac{LDP}: Using the Poseidon hash function, our circuit has $5997$~R1CS constraints. On an Ubuntu 20.04 virtual machine with $4$ virtual cores that runs on a commercial standard Laptop (Dell Latitude 7400 with an Intel~i7~8665U~CPU), proof creation -- which is typically the bottleneck for using ZKPs -- takes around $2.2$\,s when using the Groth16 proof system~\citep{groth2016size} on the  Barreto-Naehrig curve over a $254$ bit prime field (bn128), Web assembly for witness generation, and Javascript for proof generation. 
The size of the proving and verification key are around $3.4$\,MB and $3.5$\,kB, respectively. 
These sizes suggest that proof generation would also be practical on a web-based mobile application, although future research should validate this assumption. 

With an optimized tool, performance is even better: Using an optimized C++-based witness generation and a proof generation based on x86 Assembly for Intel processors (``Rapidsnark'', see~\citep{hermez2021rapidsnark}), proof generation is reduced to only $140$\,ms. 
Because to date there is no available optimized tool for proof verification, this operation still takes around $0.8$\,s in Javascript. 
Proof verification in a complex, combined survey may even reduce complexity on the surveyor's side because the complexity of \ac{SNARK} verification is not dependent on the complexity of the original computation for which the survey participant proves integrity. 
Moreover, we tested the deployment of a smart contract verifier on Ethereum, which could be used for blockchain-based, incentivized differentially private surveys and, therefore, \ac{GDPR}-compliant applications on personal data.
We measured the smart contract's deployment cost at around $1,150,000$ gas and its invocation at around $300,000$ gas.

%% file: 005_Related_work.tex
\section{Related Work}
\label{sec:Related_work}

While there is current and extensive research in the local model of DP~\cite{RR_local_DP_exterme_points, RR_local_DP_exterme_points_graphs, RR_DP_extremal, heavy_hitter_agg, binary_mech, DP_precision_Geometric} and in \acp{ZKP}~\cite{groth2006perfect, ben2013snarks, ben2019scalable}, there are only few publications that bridge both technologies. 

We identified four studies that are close to ours. R{\"u}ckel et al.~\cite{rueckel2022fairness} propose an architecture to share weights from federated learning models in a verifiable DP manner and add verifiable noise to the private weights.
However, their approach does not acknowledge that their discretization of the Laplace distribution only fulfills \emph{approximate} DP, and does not propose a bound for $\delta$ or an approach that works for high precision requirements, as they use the approximate inverse \ac{CDF} as input for the circuit.
Moreover, Tsaloli~et~al.~\cite{tsaloli2019differential} only provide a high-level motivation for using \acp{ZKP} for verifiable differential privacy, without implementation details.
Furthermore, while Kato et al.~\cite{kato2021preventing} provide details on how to create fair randomness with a related technology (secure-multiparty computation), they do not attempt to make the result of the algorithm verifiable, i.e., they cannot provide a cryptographic check of the truthful value from, e.g., an anonymous credential, before adding \ac{LDP} noise.

Lastly, Narayan et al.~\cite{narayan2015verifiable} discuss the opportunities of verifiable differential privacy, yet they provide no details of their \ac{ZKP}-based implementation. 
For instance, they do not elaborate how they consider rounding and how they achieve guarantees on the accuracy of their verifiable \emph{pure} DP proposal.
Additionally, their approach focuses on the \emph{central} model of DP instead and shows impractical performance, as it requires $2$ hours of proof generation for $32$ servers. Nonetheless, when implemented with more recent \ac{ZKP} libraries, their performance may be closer to ours as advances in proving times over the last years have been dramatic.

%% file: 006_Discussion.tex
\section{Discussion}
\label{sec:Discussion}

While adapting randomized response to \acp{ZKP} is straightforward, we investigated several approaches before successfully implementing exponentially distributed noise with \acp{ZKP}.
This section discusses the process we followed to arrive to the implementation described in Section~\ref{sec:Implementation}.

Adapting a DP mechanism that leverages exponential noise to ZKP has two significant challenges. 
Conceptually, since \acp{ZKP} can verify an arbitrary program, sampling from an exponential distribution may seem straightforward. Unfortunately, in practice, repeated operations with floats that involve rounding in classical software are challenging to implement because the range of numbers in the nominator and denominator is bounded by a large prime, and repeated rounding is costly since the complexity of the \ac{ZKP} always needs to account for the worst possible case. 
Furthermore, the propagation of the corresponding errors becomes challenging to control.
Thus, the generality of computations that \acp{ZKP} can cover well is initially limited to arithmetic operations on prime fields and their corresponding primitives such as hash functions and signatures.

The second major challenge is the inability of finite computers to fulfill the definition of \ac{DP} on the real line.
Mironov~\citep{mironov_significance_2012} was the first to demonstrate that implementing a \ac{DP} mechanism with the floating-point arithmetic of finite computers does not guarantee \ac{DP}.
Mironov proposed to solve this issue while maintaining $\varepsilon$-\ac{DP} with the Snapping mechanism~\citep{mironov_significance_2012}, and recently, Naoise et al. proposed secure random sampling~\cite{random_sampling_DP}.
However, while their output noise is discrete, we must still handle floats that \acp{ZKP} cannot process efficiently.

Therefore, we first thought of discretizing the support of the Laplace distribution (well-known to be $\varepsilon$-\ac{DP}~\citep{dwork_algorithmic_2013}) by sampling from its inverse \ac{CDF} with a finite input range $\{1, ..., d\}$, where $d$ denotes the precision Circom~\cite{circom} can handle -- similarly to Rückel et al.~\cite{rueckel2022fairness}.
However, we were not able to determine provable guarantees on $\delta$ for the approximate \ac{DP} mechanism.
Thus, we turned to the \emph{Stone-Weierstrass theorem} to approximate a polynomial so close to the Laplacian \ac{PDF} that the approximation error would be negligible.
Furthermore, because the approximated \ac{PDF} would be a polynomial itself, we thought to elegantly prove its use with \ac{ZKP}. 
We employed Bernstein polynomials~\cite{faroukibernstein2012} to approximate the \ac{PDF} in a closed interval and, subsequently, performed rejection sampling.
However, we encountered two problems: (i)~our approximation was limited to a closed interval, whereas the Laplacian \ac{PDF} has unbounded support, and (ii) the Bernstein coefficients are in general real numbers, which \ac{ZKP} cannot process efficiently, and the propagation of errors when rounding with fixed precision is again complex to handle.
Specifically, the complexity stems from the very high degree of the polynomials and the lack of homogeneity, i.e., there are many different degrees of monomials that all scale differently for a specific accuracy when multiplying inputs with a large power of $10$ and rounding afterward. 

Subsequently, we turned to the \ac{TGM}~\cite{geometric_mechanism}, which coincidentally has the advantage to provide better accuracy for count queries~\cite{garrido_i_2021}, the focus of this study.
Additionally, the truncation solved the problem of working with a closed interval (also a limitation of finite computers) by condensing the probability mass outside the interval in its lower and upper bounds.
Moreover, the support of the geometric mechanism are integers, which \acp{ZKP} can process efficiently.
Overall, \ac{TGM} adapts to finite computers while still providing pure $\varepsilon$-\ac{DP}.
However, the probabilities assigned to each integer still fall on the real line.
While we ensured these probabilities became rational numbers by carefully choosing $\varepsilon$~\cite{DP_precision_Geometric}, which a conventional computer can handle, the integers necessary to represent them were too large for the limited precision available in Circom~\cite{circom} and other libraries for implementing \acp{ZKP}, and we were unable to write a theoretical bound for $\delta$ if we approximated the real numbers with finite precision.

To cope with precision limitations and the difficulty to bound $\delta$, we then looked for simple sampling methods that provide bounds on~$\delta$, which finally led us to Dwork et al.~\cite{dwork2006noise} (see Section~\ref{sec:Implementation}).
This concluded our search, as their method for sampling exponentially distributed noise consists on repeatedly flipping unbiased coins (which is easily implemented in Circom with hashing and conversion to bit arrays), and provides a bound for $\delta$ based on the precision we can afford with Circom.

In our implementation, we used verifiable randomness co-created by the verifier (surveyor) and the prover (survey participant). 
As we noted in Section~\ref{sec:Preliminaries}, when a surveyor repeatedly conducts the query in our implementation with different challenges, they could get additional information because by the law of large numbers, the truthful query value without noise can be determined with increasing accuracy. 
Consequently, in many scenarios, it may be appropriate to use a challenge that is hard coded, derived from the surveyor institution's public key, or even derived solely from the attribute (e.g., the index of the age in the anonymous credential), such that repeated queries, even from different but colluding institutions, would not decrease the degree of plausible deniability~$\varepsilon$.
Furthermore, note that our verifiable randomized response implementation could be easily extended to flip biased coins by, e.g., generating a verifiable hash and checking whether its normalized value is lower than the desired bias.

%% file: 007_Conclusion.tex
\section{Conclusion}
\label{sec:Conclusion}

We introduce primitives for implementing verifiable differentially private polls in the local setting.
To achieve verifiability, we carefully selected \ac{DP} mechanisms for binary and numerical data and adapted their implementations to \acp{SNARK}.
Thanks to these primitives, we can achieve cryptographically verifiable survey responses while providing plausible deniability for survey participants and, in turn, not only reduce but entirely prevent bias in survey participants' answers while giving them the needed privacy guarantees.
Furthermore, note that our primitive for verifiable exponentially distributed noise allows for different aggregation queries beyond the count, as it can ingest arbitrary sensitivity -- we limited our narrative to count queries for the simplicity of the explanations.
Finally, thanks to the evaluations we performed, we conclude that practitioners can deploy our primitives with acceptable performance

We encourage practitioners to develop further primitives that can adapt to other \ac{DP} mechanisms, e.g., the exponential mechanism for categorical data~\cite{mcsherry2007mechanism}, and other randomized-response~\cite{RR_two_stage_1, RR_two_stage_2, RR_question_model, RR_forced_response, yang_wireless_2016} and \ac{LDP}~\cite{RR_local_DP_exterme_points, RR_local_DP_exterme_points_graphs, RR_DP_extremal, heavy_hitter_agg, binary_mech, DP_precision_Geometric} approaches.
Furthermore, conducting studies about how interviewees would perceive the built-in trust would allow the research community to understand how to frame polls and reassure candidates of their privacy.
Lastly, improving the precision limitations of \ac{ZKP} circuit compilers such as Circom and more literature on frameworks for bounding $\delta$ in \emph{approximate} \ac{LDP} would open the range of practical \ac{LDP} mechanisms.